\documentclass[11pt]{article}
\usepackage{amsmath,amsfonts,amssymb,graphicx,epsfig,pdflscape}
\usepackage{cite}
\usepackage[usenames]{color}
\usepackage{pstricks}
\usepackage[backref=false]{hyperref}
\hypersetup{
colorlinks=true,
citecolor=red,
linkcolor=darkblue
}
\definecolor{darkblue}{rgb}{0,0,.8}
\definecolor{red}{rgb}{1,0,0}

\numberwithin{equation}{section}

\newcommand{\nc}{\newcommand}
\nc\disp{\displaystyle}
\nc{\fh}{\hat{f}}
\nc{\muh}{\hat{\mu}}
\nc{\nuh}{\hat{\nu}}
\nc{\spos}[2]{\makebox(0,0)[#1]{$\sm{#2}$}}
\nc{\sm}[1]{{\scriptstyle #1}}
\nc{\bib}{\bibitem}
\nc{\al}{\alpha}
\nc{\g}{\gamma}
\nc{\G}{\Gamma}
\nc{\D}{\Delta}
\nc{\eps}{\epsilon}
\nc{\la}{\lambda}
\nc{\La}{\Lambda}
\nc{\var}{\varphi}
\nc{\pa}{\partial}
\nc{\nn}{\nonumber \\ }
\nc{\hf}{\frac{1}{2}}
\nc{\dz}{\frac{dz}{2\pi i}}
\nc{\bin}[2]{\left(\!\!\!\begin{array}{c} {#1}\\ {#2} \end{array}\!\!\!\right)}
\nc{\be}{\begin{equation}}
\nc{\ee}{\end{equation}}
\nc{\bea}{\begin{eqnarray}}
\nc{\eea}{\end{eqnarray}}
\nc{\bra}[1]{\langle {#1}|}
\nc{\ket}[1]{|{#1}\rangle}
\nc{\ketw}[1]{({#1})^{\phantom{a}}_{{\cal W}}}
\nc{\chit}{\raisebox{0.25ex}{$\chi$}}
\nc{\chih}{\raisebox{0.25ex}{$\widehat\chi$}}
\nc{\ch}{{\rm ch}}
\nc{\chh}{\widehat{\mathrm{ch}}}
\nc{\Dti}{\tilde{\D}}
\nc{\Db}{\mbox{\boldmath $D$}}
\nc{\Hb}{\mbox{\boldmath $H$}}
\nc{\Ib}{\mbox{\boldmath $I$}}
\nc{\qb}{\bar{q}}
\nc{\kb}{\bar{k}}
\nc{\Fb}{\bar{F}}
\nc{\ph}{\hat{p}}
\nc{\Ph}{\hat{P}}
\nc{\qh}{\hat{q}}
\nc{\rh}{\hat{r}}
\nc{\sh}{\hat{s}}
\nc{\Mh}{\hat{M}}
\nc{\Vh}{\hat{V}}
\nc{\Vch}{\hat{\mathcal{V}}}
\nc{\Ac}{\mathcal{A}}
\nc{\Bc}{\mathcal{B}}
\nc{\Cc}{\mathcal{C}}
\nc{\Dc}{\mathcal{D}}
\nc{\Ec}{\mathcal{E}}
\nc{\Fc}{\mathcal{F}}
\nc{\Hc}{\mathcal{H}}
\nc{\Ic}{\mathcal{I}}
\nc{\Jc}{\mathcal{J}}
\nc{\Oc}{\mathcal{O}}
\nc{\Qc}{\mathcal{Q}}
\nc{\Rc}{\mathcal{R}}
\nc{\Vc}{\mathcal{V}}
\nc{\Wc}{\mathcal{W}}
\nc{\Xc}{\mathcal{X}}
\nc{\Yc}{\mathcal{Y}}
\nc{\Zc}{\mathcal{Z}}
\nc{\fus}{\mbox{}\,\hat\otimes\,\mbox{}}
\def\vvdots{\mathinner{\mkern1mu\raise1pt\vbox{\kern7pt\hbox{.}}\mkern2mu
  \raise4pt\hbox{.}\mkern2mu\raise7pt\hbox{.}\mkern1mu}}
\nc{\gauss}[2]{\left[\!\!\begin{array}{c} {#1}\\ {#2} \end{array}\!\!\right]}
\nc{\sbin}[2]{\left\{\!\!\!\begin{array}{c} {#1}\\ {#2} 
\end{array}\!\!\!\right\}}
\nc{\sbinlr}[2]{\Big\langle\!\!\begin{array}{c} {#1}\\ {#2} 
\end{array}\!\!\Big\rangle}
\nc{\bino}[2]{\left(\!\!\begin{array}{c} {#1}\\ {#2} \end{array}\!\!\right)}
\def\half {\mbox{$\textstyle \frac{1}{2}$}}

\definecolor{lightblue}{rgb}{.61,.61,1}
\definecolor{midblue}{rgb}{.7,.7,1}
\definecolor{lightlightblue}{rgb}{.85,.85,1}
\definecolor{lightestblue}{rgb}{.96,.96,1}
\definecolor{lightpurple}{rgb}{1,.65,1}
\nc{\R}{{\cal R}}
\nc{\dkk}{\delta_{j,\{k,k'\}}^{(2)}}
\nc{\drr}{\delta_{j,\{r,r'\}}^{(2)}}
\nc{\ddkk}{\delta_{j,\{k,k'\}}^{(4)}}
\nc{\dddkk}{\delta_{j,\{k,k'\}}^{(8)}}
\nc{\dnn}{\delta_{j,\{n,n'\}}^{(2)}}
\nc{\ddnn}{\delta_{j,\{n,n'\}}^{(4)}}
\nc{\dddnn}{\delta_{j,\{n,n'\}}^{(8)}}

\definecolor{pink}{rgb}{1,.65,.65}
\def\equiveq{=}

\begin{document}

\topmargin -5mm
\oddsidemargin 5mm

\setcounter{page}{1}

\vspace{8mm}
\begin{center}
{\LARGE {\bf Coset construction of logarithmic minimal models:}}
\\[.25cm]
{\LARGE {\bf branching rules and branching functions}}

\vspace{8mm}
{\Large Paul A. Pearce$^\ast$\, and\; J{\o}rgen Rasmussen$^\dagger$}
\\[.4cm]
{\em {}$^\ast$Department of Mathematics and Statistics, University of Melbourne}\\
{\em Parkville, Victoria 3010, Australia}
\\[.4cm]
{\em {}$^\dagger$School of Mathematics and Physics, University of Queensland}\\
{\em St Lucia, Brisbane, Queensland 4072, Australia}
\\[.4cm]
{\tt p.pearce\,@\,ms.unimelb.edu.au}
\qquad
{\tt j.rasmussen\,@\,uq.edu.au}
\end{center}

\vspace{12mm}
\centerline{{\bf{Abstract}}}
\vskip.4cm
\noindent
Working in the Virasoro picture, it is argued that the logarithmic minimal models 
$\mathcal{LM}(p,p')=\mathcal{LM}(p,p';1)$ can be extended to an infinite hierarchy of logarithmic 
conformal field theories $\mathcal{LM}(p,p';n)$ at higher fusion levels $n\in\mathbb{N}$. 
From the lattice, these theories are constructed 
by fusing together $n\times n$ elementary faces of the appropriate $\mathcal{LM}(p,p')$ models. 
It is further argued that all of these logarithmic theories are realized as diagonal cosets
${(A_1^{(1)})_k\oplus(A_1^{(1)})_n}/{(A_1^{(1)})_{k+n}}$ where $n$ is the integer fusion level 
and $k=\frac{np}{p'-p}-2$ is a fractional 
level. These cosets mirror the cosets of the higher fusion level minimal models of the form $\mathcal{M}(M,M';n)$, but are associated with 
certain reducible representations. We present explicit branching rules for characters in the form of 
multiplication formulas arising in the logarithmic 
limit of the usual Goddard-Kent-Olive coset construction of the non-unitary minimal models $\mathcal{M}(M,M';n)$. 
The limiting branching functions play the role of Kac characters for the $\mathcal{LM}(p,p';n)$ theories.
\renewcommand{\thefootnote}{\arabic{footnote}}
\setcounter{footnote}{0}

\tableofcontents

\newpage

\section{Introduction}

Logarithmic conformal field theories (CFTs)~\cite{RS1992,Gur1993} are relevant in describing a variety of physical systems with 
nonlocal degrees of freedom.
Arguably, the simplest and most studied logarithmic CFTs are logarithmic extensions~\cite{Flohr9509,GK9606,Flohr9605} of the 
minimal models ${\cal M}(M,M')$~\cite{BPZ1984}. 
These logarithmic theories came into sharper focus as a result of their 
realization as a family of exactly solvable lattice models ${\cal LM}(p,p')$~\cite{PRZ0607}. 
For the minimal models, $M,M'$ must satisfy $\mathrm{gcd}(M,M')=1$ with \mbox{$2\le M<M'$} whereas these constraints are 
relaxed to $\mathrm{gcd}(p,p')=1$ with $1\leq p<p'$ for the logarithmic counterparts.
The first two members of the ${\cal LM}(p,p')$ family are critical dense polymers ${\cal LM}(1,2)$ and critical percolation 
${\cal LM}(2,3)$. Many of the properties of the logarithmic minimal models  ${\cal LM}(p,p')$ mirror the properties of minimal 
models  ${\cal M}(M,M')$.

It is well known that the rational minimal models ${\cal M}(M,M')={\cal M}(M,M';1)$ admit a 
generalization~\cite{DJMO86,DJMO87,DJKMO87} 
${\cal M}(M,M';n)$ to higher integer fusion levels $n>1$ with $\mathrm{gcd}(M,\frac{M'-M}{n})=1$ and $2\le M<M'$. From within CFT, 
these theories are realized as $su(2)$ cosets via the construction of Goddard-Kent-Olive (GKO)~\cite{GKO85,GKO86}. 
From the lattice, the exactly solvable models associated with these rational CFTs are constructed by fusing together 
$n\times n$ elementary faces of the RSOS models~\cite{ABF,FB}.
In this paper, we introduce generalizations ${\cal LM}(p,p';n)$ of the logarithmic minimal models ${\cal LM}(p,p')$ $={\cal LM}(p,p';1)$ 
to higher integer fusion levels $n>1$ with $\mathrm{gcd}(p,\frac{p'-p}{n})=1$ and $1\le p<p'$. In analogy to the rational case, these new exactly solvable models are constructed on the lattice by fusing together $n\times n$ elementary faces of appropriate $\mathcal{LM}(p,p')$ models. For $n=2$, this yields the family of logarithmic superconformal 
minimal models~\cite{PRT} $\mathcal{LSM}(p,p')=\mathcal{LM}(p,p';2)$ which includes superconformal polymers $\mathcal{LSM}(1,3)$
and superconformal percolation $\mathcal{LSM}(2,4)$ as its first members. The general 
${\cal LM}(p,p';n)$ CFTs are expected to be logarithmic in the sense that they admit indecomposable representations
of rank greater than $1$.
This is known for the case $n=1$ and will be established for $n=2$ in \cite{PRT}.

Perhaps more importantly, this paper initiates a study of the coset construction of the logarithmic $su(2)$ coset models 
${\cal LM}(p,p';n)$ in the Virasoro picture\footnote{We use the term {\em Virasoro picture} to indicate that
the conformal symmetry algebra is just the Virasoro algebra. 
Similarly, we refer to the {\em ${\cal W}$-extended picture} when the conformal symmetry algebra is enlarged to the 
${\cal W}_{p,p'}$ algebra of~\cite{FGST0606}.}.
Coset constructions use knowledge of affine current algebra building blocks to gain insight into large families of CFTs. Due to the broad applicability of these constructions, they have been studied 
extensively~\cite{BG1987,GK88,MP90,MW90,BMP91,MSW9110,DJ93,GW9407,HR95,Sch9508,MRW0007,FS0111,Ish0111,FFRS0309,GG1011,Ahn1106} 
and used in various classification schemes of rational 
CFTs~\cite{BS9210,BEHHH9406,DiFMS} placing due emphasis on the Lie algebraic building blocks. 
In particular, while the Kac determinant~\cite{Kac1979,FF1984} gives necessary 
conditions for a minimal model representation to be unitary~\cite{FQS}, the existence of these representations was provided 
by coset constructions~\cite{GKO86}.

A myriad of coset constructions exist, but here we only consider diagonal cosets of the form
\be
 \mbox{COSET}(k,n):\quad \frac{(A_1^{(1)})_k\oplus(A_1^{(1)})_n}{(A_1^{(1)})_{k+n}},
 \qquad k=\frac{\ph}{\ph'}-2,\qquad \mathrm{gcd}(\ph,\ph')=1,\qquad n,\ph,\ph'\in\mathbb{N}
\label{cosetAAA}
\ee
where  $\mathbb{N}$ denotes the set of positive integers and the subscripts on the 
affine $su(2)$ current algebra $A_1^{(1)}$ denote the respective levels $k$, $n$ and $k+n$.
We assume that the fusion level $n$ is a positive integer and that $k$ is a fractional level with $k>-2$. 
The level $k$ is called admissible if $\ph\geq2$, but in the logarithmic setting we move beyond admissible levels and allow $\ph=1$. 
An integral part of the coset construction (\ref{cosetAAA}) is the set of {\em branching rules} determining the decompositions of products of affine characters back into affine characters. 
The coefficients in these decompositions are called {\em branching functions}. 
They are characters of the coset model. Although the branching functions are properly associated with CFT and critical lattice models, it is of interest to observe that they also appear in the study of off-critical lattice models (up to their leading powers of $q$) as the one-dimensional configurational sums in Baxter's 
corner transfer matrix 
method. This is true for both off-critical rational minimal models~\cite{DJMO87,DJKMO87} and off-critical logarithmic models~\cite{PS1207}.

By considering {\em admissible representations}~\cite{KW88,KW88b} 
of the affine current algebras, the construction (\ref{cosetAAA}) gives rise
to the infinite family of minimal $su(2)$ coset models~\cite{ACT91,BMSW9702} 
\be
 {\cal M}(M,M';n)\simeq \mbox{COSET}\Big(\frac{nM}{M'-M}-2,n\Big),\quad \mathrm{gcd}\Big(M,\frac{M'-M}{n}\Big)=1,\quad
   2\leq M<M',\quad n,M,M'\in\mathbb{N}
\label{MMMn}
\ee
The usual minimal models ${\cal M}(M,M')={\cal M}(M,M';1)$ correspond to $n=1$ 
while the superconformal minimal models ${\cal SM}(M,M')={\cal M}(M,M';2)$ are given by $n=2$. 

To obtain the logarithmic branching rules and branching functions of ${\cal LM}(p,p';n)$, we apply a {\em logarithmic limit\/} 
to the GKO coset (\ref{cosetAAA}) of the minimal $su(2)$ coset models ${\cal M}(M,M';n)$.
Formally, this general procedure is described by
\be
 {\cal LM}(p,p';n)=\lim_{M,M'\to\infty\atop M/M'\to p/p'}\!{\cal M}(M,M';n),\qquad \mathrm{gcd}(p,p')=1,\quad 1\leq p<p',\quad p,p'\in\mathbb{N}
 \label{logLimit}
\ee
where the limit is taken through a sequence of $M,M'$ values satisfying 
\be
\mathrm{gcd}\Big(M,\frac{M'-M}{n}\Big)=1,\quad
   2\leq M<M',\quad M,M'\in\mathbb{N}
\label{pplimits}
\ee
This limiting procedure was developed~\cite{Ras0405,Ras0406} as a means to describe and 
analyze non-rational logarithmic 
CFTs using knowledge of a suitable infinite series of rational CFTs.
The logarithmic limit has been successfully applied to gain insight into
the $\mathcal{LM}(p,p';1)$ models at criticality~\cite{Ras0405,Ras0406} 
and to construct off-critical loagrithmic minimal models~\cite{PS1207}.
In particular, as recalled in Section~\ref{SecKac}, the infinite family of Kac characters of ${\cal LM}(p,p';1)$ arise as limits of 
non-unitary minimal Virasoro characters.
The logarithmic limit has also been used to compute correlation functions and anomalies in gravity duals to 
logarithmic CFTs~\cite{GJZ1010,GRRZ1302}.
There are some subtleties~\cite{Ras0405,Ras0406} regarding the emergence of Jordan-block structures
and indecomposability properties of the representations in the logarithimic limit. Nevertheless, it appears that the logarithmic limit is robust 
and independent of the choice of the sequence of $M,M'$ values (\ref{pplimits}) 
when its application is restricted to the spectra of chiral theories as in this paper. The logarithmic limit may require modification to apply to bulk theories on a torus.

In this paper, we focus on the branching rules and branching functions arising from the coset construction (\ref{MMMn}) in the logarithmic limit (\ref{logLimit}). 
In a forthcoming paper~\cite{PR2013b}, we adopt a constructive approach by interpreting the logarithmic limit
as a coset of the form (\ref{cosetAAA}) in its own right
\be
 {\cal LM}(p,p';n)\simeq \mbox{COSET}\Big(\frac{np}{p'-p}-2,n\Big),\qquad \mathrm{gcd}\Big(p,\frac{p'-p}{n}\Big)=1,\quad 
    1\leq p<p',\quad n,p,p'\in\mathbb{N}
\label{Clog}
\ee
but based primarily on {\em reducible} representations of the affine current algebras.
Specifically, we have in mind the affine Kac characters (\ref{affKac}).
We refer to them as {\em affine Kac characters} since,
for $n=1$, they are reminiscent of the Kac characters of the 
models $\mathcal{LM}(p,p';1)$. In the corresponding branching rules,
the Kac characters of the coset theory appear as branching functions. 

Alternatively, in the 
${\cal W}$-extended picture, the $\mathcal{LM}(p,p')$ models can be viewed~\cite{FGST0606} with an extended ${\cal W}_{p,p'}$ chiral symmetry algebra instead of just the Virasoro algebra. 
From the lattice perspective, the 
corresponding ${\cal W}$-extended CFTs $\mathcal{WLM}(p,p')$ are described~\cite{PRR0803,RP0804,Ras0805} by the 
{\em same} $\mathcal{LM}(p,p')$ lattice models 
but with a special class of restricted boundary conditions which, in the continuum scaling limit, respect the $\cal W$-extended symmetry. 
Construction of other cosets of logarithmic CFTs with ${\cal W}$-symmetry first appeared in~\cite{Sem0710,Sem1109}, 
whereas the first hints 
of the existence of a $\cal W$ diagonal coset construction of the ${\cal WLM}(p,p')$ models, of an analogous form to 
(\ref{cosetAAA}), appeared
in~\cite{PR1010}. Specifically, it was found that the Grothendieck ring associated with ${\cal W}$-projective representations
leads to a Verlinde-like formula involving coset graphs $A_p^{(2)}\otimes A_{p'}^{(2)}/\mathbb{Z}_2$ based on twisted affine 
Dynkin diagrams, and that modular invariants of the models are encoded by the same coset graphs. Concrete coset constructions
of the logarithmic minimal models, however, have remained elusive.

The layout of this paper is as follows.
In Section~\ref{SecCoset}, we recall the rational GKO construction of the 
minimal $su(2)$ coset models ${\cal M}(M,M';n)$ 
using admissible representations and review 
the corresponding branching rules and branching functions.
In Section~\ref{SecLimit}, we take the logarithmic limit of the minimal branching rules. 
The ensuing logarithmic branching functions are identified as 
(Virasoro) Kac characters of the logarithmic ${\cal LM}(p,p';n)$ models. 
Section~\ref{SecConcl} contains some concluding remarks.

\section{Coset construction}
\label{SecCoset}

Let $h$ denote an affine current subalgebra of the affine current algebra $g$,
and let $L^{h}_n$ and $L^g_n$ denote the generators of the corresponding
Segal-Sugawara constructions of central charge $c^g$ and $c^h$, respectively. 
In the seminal papers~\cite{GKO85,GKO86} by Goddard, Kent and Olive, it was found that
\be
 L_n^{g/h}:=L_n^g-L_n^{h},\qquad n\in\mathbb{Z}
\ee
generate a Virasoro algebra of central charge
\be
 c^{g/h}=c^g-c^{h}
\ee
and that this {\em GKO coset Virasoro algebra} commutes with the Virasoro
algebra generated by $L_n^{h}$, that is,
\be
 [L_n^{g/h},L_m^{h}]=0,\qquad n,m\in\mathbb{Z}
\ee
More generally, one has
\be
 [L_n^{g/h},J_m^{a}]=0,\qquad J_m^{a}\in h
\ee
In a decomposition of a $g$-module $\mathcal{M}_\la^g$ in terms of
$h$-modules $\mathcal{M}_{\mu}^h$
\be
 \mathcal{M}_\la^g\simeq\bigoplus_{\mu}B_\la^{\mu}\otimes\mathcal{M}_{\mu}^h
\ee
it is therefore natural to interpret $B_\la^{\mu}$ as a module over the coset Virasoro
algebra generated by $L_n^{g/h}$. At the level of characters, one thus has the
{\em branching rules}
\be
 \chih[\mathcal{M}_\la^g](q,z)=\sum_{\mu}\chit[B_\la^{\mu}](q)\,\chih[\mathcal{M}_{\mu}^h](q,z)
\label{chiMBM}
\ee
where the {\em branching functions} $\chit[B_\la^{\mu}](q)$ are interpreted as
Virasoro characters of the {\em coset conformal field theory} described by $g/h$.
Certain linear combinations of these branching functions may subsequently form 
extended characters, such as superconformal characters, in the coset model.

Here we are interested in the {\em coset construction} (\ref{cosetAAA}) where
$h=(A_1^{(1)})_{k+n}$ is the diagonal subalgebra of $g=(A_1^{(1)})_k\oplus(A_1^{(1)})_n$.
The central charge of the coset Virasoro algebra is thus given by
\be
 c=c_k+c_n-c_{k+n}
  =\frac{3kn(k+n+4)}{(k+2)(n+2)(k+n+2)}
\ee
where the central charge of the affine current algebra $(A_1^{(1)})_k$ is given by
\be
 c_k=\frac{3k}{k+2}
\label{ck}
\ee
A primary objective of this work is to examine the branching rules and branching functions based on
certain classes of $(A_1^{(1)})_k$ and $(A_1^{(1)})_n$
characters for $n,\ph,\ph'\in\mathbb{N}$, where the fractional level is $k=\frac{\ph}{\ph'}-2$.
Hence, we are not concerned with so-called 
field identifications~\cite{Gepner 1989,AW1990,SY1990,FSS9509}, 
for example, which are applied when
a branching function occurs in more than one branching rule in a given coset construction.

\subsection{Admissible affine characters}

The character of an $(A_1^{(1)})_k$ module ${\cal L}$ is defined by
\be
 \chih[{\cal L}](q,z):=\mathrm{Tr}_{{\cal L}}^{\phantom{V}}\,q^{L_0-\frac{c_k}{24}}z^{J_0^3}
\ee
where $L_0$ and $J_0^3$ are 
zero modes of the affine current algebra $(A_1^{(1)})_k$ and the modular nome $q$ satisfies $|q|<1$.
To describe such characters, we introduce the standard Jacobi-Riemann theta functions
\be
 \vartheta_{n,m}(q,z):=\sum_{\ell\in\mathbb{Z}+\frac{n}{2m}}q^{m\ell^2}z^{m\ell},\qquad 
  n\in\mathbb{Z},\quad m\in\mathbb{N}
\ee

In this paper, we are interested in the affine current algebra $(A_1^{(1)})_k$ at fractional (or integer) level and will focus on
the cases for which the shifted level 
\be
 t:=k+2>0
\ee 
is positive. From (\ref{cosetAAA}), we thus have the parameterization
\be
 t=\frac{\ph}{\ph'},\qquad \mathrm{gcd}(\ph,\ph')=1,\qquad \ph,\ph'\in\mathbb{N}
\label{t}
\ee
where the notation $\ph,\ph'$ is used to distinguish these parameters from the
ones in ${\cal LM}(p,p')$. Such a level is said to be {\em admissible} if
\be
 \ph\geq2
\label{p2}
\ee
in which case the associated weights $j_{r,s}$ of the form
\be
 2j_{r,s}+1=r-s t,\qquad 
1\leq r\leq \ph-1,\qquad0\leq s\leq \ph'-1
\label{j}
\ee
are called {\em admissible weights}.
In most applications, one works with the corresponding irreducible highest-weight representations,
also known as {\em admissible representations}~\cite{KW88,KW88b}. Their characters 
are called {\em admissible characters} and are given by
\be
\begin{array}{rcl}
 \chh_{r,s}^{\ph,\ph'}(q,z)
 &\!\!=\!\!&\disp\frac{\vartheta_{\la_{r,s}^+,\ph\ph'}(q,z^{\frac{1}{\ph'}})
   -\vartheta_{\la_{r,s}^-,\ph\ph'}(q,z^{\frac{1}{\ph'}})}{\eta(q,z)},\qquad 
 \la_{r,s}^\pm=\pm r\ph'-\ph s\\[10pt]
 &\!\!=\!\!&\disp\frac{1}{\eta(q,z)}\sum_{\ell\in\mathbb{Z}}
  q^{\ph\ph'(\ell+\frac{\la_{r,s}^+}{2\ph\ph'})^2} z^{\ph(\ell+\frac{\la_{r,s}^+}{2\ph\ph'})}
  \big(1-q^{-r(2\ell \ph'-s)}\,z^{-r}\big)
 \end{array}
\label{chadm}
\ee
where the `affine' Dedekind eta function $\eta(q,z)$ is defined by
\be
 \eta(q,z):=q^{\frac{1}{8}}z^{\frac{1}{2}}\prod_{n=1}^\infty (1-q^nz)(1-q^{n-1}z^{-1})(1-q^n)=\vartheta_{1,2}(q,z)-\vartheta_{-1,2}(q,z)
\ee

In the coset expression (\ref{cosetAAA}), one of the constituent current algebras
has the integer level $n$,
in which case the admissible characters can be written as
\be
 \widehat{\ch}_{\rho,0}^{n+2,1}(q,z)=\frac{1}{\eta(q,z)}\sum_{\ell\in\mathbb{Z}}
  q^{(n+2)(\ell+\frac{\rho}{2(n+2)})^2}z^{(n+2)\ell+\frac{\rho}{2}}\big(1-q^{-2\ell \rho}z^{-\rho}\big),
   \qquad \rho=1,2,\ldots,n+1
\ee
For $n=1$, in particular, there are two such characters, and they admit the simplifications
\be
 \widehat{\ch}_{\rho,0}^{3,1}(q,z)=\frac{\vartheta_{\rho,3}(q,z)
   -\vartheta_{-\rho,3}(q,z)}{\eta(q,z)}
 =\frac{\vartheta_{\rho-1,1}(q,z)}{\eta(q)}=\frac{1}{\eta(q)}\sum_{\ell\in\mathbb{Z}+\frac{\rho-1}{2}}q^{\ell^2}z^\ell,
  \qquad \rho=1,2
\ee
where the usual Dedekind eta function $\eta(q)$ is defined in terms of the Euler product $(q)_\infty$ by
\be
 \eta(q):=q^{\frac{1}{24}}(q)_\infty,\qquad 
  (q)_\infty:=\prod_{j=1}^\infty(1-q^j)
\ee

\subsection{Rational branching rules for minimal $su(2)$ coset models}
\label{SecRational}

Here we recall the coset construction of the minimal $su(2)$ coset models (\ref{MMMn})
and the structure of the corresponding rational branching rules and branching functions~\cite{ACT91,BMSW9702}.
The cosets are of the form (\ref{cosetAAA}) and based on admissible characters of the three affine current algebras
$(A_1^{(1)})_{k}$, $(A_1^{(1)})_{n}$ and $(A_1^{(1)})_{k+n}$.

Recalling the parameterization $k+2=\frac{\ph}{\ph'}$, 
the minimal $su(2)$ coset model (\ref{MMMn})
given by the coset construction (\ref{cosetAAA}) is
\be
 {\cal M}(M,M';n),\qquad M=\ph,\quad M'=\ph+n\ph'
\label{MMM}
\ee
Its central charge is given by 
\be
 c^{M,M'\!;n}=\frac{3n}{n+2}\Big(1-\frac{2(n+2)(M-M')^2}{n^2MM'}\Big)
\label{cnMM}
\ee
Since the underlying affine characters are admissible, it follows that (\ref{MMM}) implies the familiar constraints
\be
 \mathrm{gcd}\big(M,\frac{M'-M}{n}\big)=1,\qquad 2\leq M<M'
\ee

The rational branching functions of the minimal $su(2)$ coset models are expressible in terms of the
string functions~\cite{KP84,JM84,GQ87,DQ90,HNY90} of $\mathbb{Z}_n$ 
parafermions with central charge $c=\frac{2n-2}{n+2}$. For the fundamental domain
\be
 0\leq m\leq\ell\leq n,\qquad \ell-m\in2\mathbb{Z},\qquad m,\ell=0,1,\ldots,n,\qquad 
n\in {\mathbb N}
\label{stringdomain}
\ee 
these string functions are given by
\bea
 c_m^\ell(q)&=&\frac{q^{-\frac{1}{24}\frac{2n-2}{n+2}+\frac{\ell(\ell+2)}{4(n+2)}-\frac{m^2}{4n}}}{(q)_\infty^3}
  \sum_{i,j=0}^\infty(-1)^{i+j}q^{ij(n+1)+\frac{1}{2}i(i+1)+\frac{1}{2}j(j+1)}\nn
&&\qquad\quad\mbox{}\times\big[q^{\frac{i}{2}(\ell+m)+\frac{j}{2}(\ell-m)}-q^{n-\ell+1+\frac{i}{2}(2n+2-\ell-m)+\frac{j}{2}(2n+2-\ell+m)}\big]
\eea
where the dependence on $n$ has been suppressed.
The notation for the string functions should not be confused with the notation for the central charges.
The fundamental domain (\ref{stringdomain}) of definition of the string functions is extended to the domain
\be
 \ell=0,1,\ldots,n,
 \qquad m\in\mathbb{Z},\qquad n\in\mathbb N
\label{elldomain}
\ee
by setting $c_m^\ell(q)=0$ for $\ell-m\notin 2\mathbb{Z}$ and using the symmetries
\bea
 c_m^\ell(q)=c_{-m}^\ell(q)=c_{n-m}^{n-\ell}(q)=c_{m+2n}^\ell(q)\label{stringsym}
\eea
so that $c_m^\ell(q)$ is even and periodic in $m$ with period $2n$. 

In terms of string functions, the {\em minimal branching functions} are given by~\cite{ACT91,BMSW9702}
\bea
 b_{r,s;\ell}^{M,M'\!;n}(q)&=&q^{\Delta_{r,s}^{M,M'\!;n}-\frac{c^{M,M'\!;n}}{24}+\frac{n-1}{12(n+2)}}\nn
 &\times& \sum_{0\le m\le n/2\atop m\equiveq\ell/2\;\text{mod 1}} \!\!\!\!\!c_{2m}^\ell(q) 
\bigg[\!\!\sum_{j\in\mathbb{Z}\atop m_{r-s}(j)\equiveq \pm m\;\text{mod}\;n}\!\!\!\!\!\!\!\!\!\!\!\!\! q^{{j\over n}(jM M'+rM'-sM)}
-\!\!\!\!\!\!\sum_{j\in\mathbb{Z}\atop m_{r+s}(j)\equiveq \pm m\;\text{mod}\;n}\!\!\!\!\!\!\!\!\!\!\!\!\! q^{{1\over n}(jM'+s)(jM+r)}\bigg]
\label{b}
\eea
where 
\be
 \D_{r,s}^{M,M'\!;n}=\frac{(rM'-sM)^2-(M'-M)^2}{4nM M'},\qquad 1\le r\le M\!-\!1,\ \ 1\le s\le M'\!-\!1
\label{DrsMMn}
\ee
$r,s$ and $\ell$ are bounded as in (\ref{DrsMMn}) and (\ref{elldomain}), respectively, and
\be
 m_a(j):=a/2+jM'
\ee
The first sum in (\ref{b}) runs over integers $m$ if $\ell$ is even and half odd integers 
$m$ if $\ell$ is odd.
The restriction in the sum over $j$ indicates that the sum is only over those values of $j$ for which 
$m_a(j)=\pm m\ \mbox{mod $n$}$.
The last term in the leading power of $q$ in (\ref{b}) is associated with the central charge of  $\mathbb{Z}_n$ 
parafermions.

The branching functions $b_{r,s;\ell}^{\,M,M'\!;n}(q)$ and their associated conformal weights 
$\Delta_{r,s;\ell}^{M,M'\!;n}$ are restricted to the checkerboard
\be
r+s\equiveq\ell \mbox{\ \ mod\;2},\qquad \ell=0,1,\ldots,n
\label{checkerboard}
\ee
We have not been able to find general expressions for the minimal coset conformal weights of these branching functions in the literature. 
However, an analysis of (\ref{b}) yields the explicit formula
\be
 \D_{r,s,\ell}^{M,M'\!;n}= \D_{r,s}^{M,M'\!;n}+\D_{r-s}^{\ell;n}+\mbox{Max}[\half(\ell\!+\!2\!-\!r\!-\!s),0]+\mbox{Max}[\half\big(\ell'\!+\!2\!-\!(M\!-r)\!-\!(M'\!-\!s)\big),0]
 \label{cosetDelta}
\ee
where, setting $m'=m$ mod $2n$, 
\be
\hspace{-6pt}\Delta_m^{\ell;n}=\mbox{Max}[\Delta(m',\ell,n),\Delta(2n\!-\!m',\ell,n),\Delta(n\!-\!m',n\!-\!\ell,n)],\qquad
\Delta(m,\ell,n)={\ell(\ell+2)\over 4(n+2)}-{m^2\over 4n}
\label{stringConfWts}
\ee
is the conformal weight of the string function $c_m^\ell(q)$ folded into the fundamental domain (\ref{stringdomain}) and
\be
\ell'=\begin{cases}
\ell,&\mbox{$\frac{M'-M}{n}$ even}\\[2pt]
n-\ell,&\mbox{$\frac{M'-M}{n}$ odd}
\end{cases}
\ee
The third term on the right side of (\ref{cosetDelta}) arises when there is cancellation in the leading power of $q$ between the two sums in square brackets in (\ref{b}). 
This term only gives a nonzero contribution for $r+s\le \ell\le n$. 
The fourth term is the counterpart of the third term under the Kac table symmetry. 
It only contributes for $r+s\ge M+M'-\ell'\ge M+M'-n$. 
The conformal weights (\ref{cosetDelta}) are thus conveniently organized into $n+1$ layered Kac tables each displaying the 
checkerboard pattern (\ref{checkerboard}) and satisfying the Kac table symmetry
\be
 \D_{r,s;\ell}^{M,M'\!;n}=\Delta^{M,M';n}_{M-r,M'-s;\ell'}
\ee
For $n=1$, the two checkerboards can be combined to form the usual single layer Kac table with the usual Kac table symmetry. 
In this case, only the first of the four terms in (\ref{cosetDelta}) survives. 

The branching functions (\ref{b}) satisfy~\cite{ACT91,BMSW9702} the {\em minimal branching rules}
\be
 \chh_{r,s}^{\ph,\ph'}(q,z)\,\chh_{\rho,0}^{n+2,1}(q,z)
  =\sum_{\mbox{\scriptsize$\sigma=1$}\atop \mbox{\scriptsize $\sigma\equiveq r\!+\!\ell$ mod 2}}^{\ph+n\ph'-1}
    b_{r,\sigma;\ell}^{\,\ph,\ph+n\ph'\!;n}(q)\,\chh_{\sigma,s}^{\ph+n\ph'\!,\ph'}\!(q,z)
\label{branchR}
\ee
relating admissible characters of the affine current algebras $(A_1^{(1)})_k$, $(A_1^{(1)})_n$ and $(A_1^{(1)})_{k+n}$ where
\be
 \ell=
\begin{cases}
 n+1-\rho,&\mbox{$s$ odd}
\\
 \rho-1,&\mbox{$s$ even}
\end{cases}
\label{ell}
\ee

\section{Logarithmic limit}
\label{SecLimit}

Logarithmic limits of rational CFTs were  
developed in~\cite{Ras0405,Ras0406} as a means to describe and analyze logarithmic CFTs.
In this way, the Kac characters of the logarithmic minimal models arise as limits of minimal Virasoro characters.
Here we apply the limiting procedure to the branching rules and branching functions of the GKO construction discussed in 
Section~\ref{SecRational}.
Before discussing the logarithmic limit of affine characters in Section~\ref{SecAffineKac}, 
we recall in Section~\ref{SecKac}
how the Kac characters of the logarithmic minimal model ${\cal LM}(p,p')$ are obtained by taking 
the appropriate logarithmic limit of irreducible minimal Virasoro characters.

\subsection{Kac characters}
\label{SecKac}

The irreducible characters associated with the rational Kac table of the minimal model ${\cal M}(M,M')$ are given by
\bea
\begin{array}{rcl}
 \ch_{r,s}^{M,M'}\!(q)&\!\!=\!\!&\disp\frac{\vartheta_{\La_{r,s}^+,MM'}(q)
   -\vartheta_{\La_{r,s}^-,MM'}(q)}{\eta(q)},\qquad 
 \La_{r,s}^\pm=\pm rM'-Ms\\[12pt]
 &\!\!=\!\!&\disp {q^{-{c^{M,M'\!;1}\over 24}+\Delta_{r,s}^{M,M'\!;1}}\over (q)_\infty}
\sum_{k=-\infty}^\infty \!\!\!\big[q^{k(k M M'+rM'-sM)}\!-\!q^{(kM+r)(kM'+s)}\big]
\end{array}
\eea
where
\be
 1\leq r\leq M-1,\qquad 1\leq s\leq M'-1
\ee
and where the specialized Jacobi-Riemann theta functions are defined as
\be
 \vartheta_{n,m}(q):=\vartheta_{n,m}(q,1)=\sum_{\ell\in\mathbb{Z}+\frac{n}{2m}}q^{m\ell^2},\qquad 
  n,m\in\mathbb{Z}
\label{vartspec}
\ee
Using the fact that $|q|<1$, it follows that
\be
 \lim_{M,M'\to\infty\atop M/M'\to p/p'}\ch_{r,s}^{M,M'}(q)=\chit_{r,s}^{p,p'}(q)
\ee
where the limit is taken through a sequence as in (\ref{pplimits}) and where
\be
\begin{array}{c}
 \chit_{r,s}^{p,p'}(q)=\disp{\frac{q^{\D_{r,s}^{p,p'}-\frac{c^{p,p'}}{24}}}{(q)_\infty}(1-q^{rs})}
\\[.5cm]
 c^{p,p'}=\disp{1-\frac{6(p-p')^2}{pp'}},\qquad \D_{r,s}^{p,p'}=\disp{\frac{(rp'-ps)^2-(p'-p)^2}{4pp'}},\qquad r,s\in\mathbb{N}
\end{array}
\label{Kacchar}
\ee
These formulas for the logarithmic central charge and conformal weights are exactly as in the minimal formulas (\ref{cnMM}) and (\ref{DrsMMn}), for $n=1$, but with an  extended range for $r,s$. 
The character $\chit_{r,s}^{p,p'}(q)$ is recognized as a {\em Kac character}~\cite{PRZ0607} of the logarithmic minimal model 
${\cal LM}(p,p')$ and is the 
character of a so-called Kac module~\cite{Ras1012,BGT1102}. These characters are associated with an
{\em infinitely} extended Kac table as the Kac labels $r$ and $s$ are unbounded from above.

\subsection{Affine Kac characters}
\label{SecAffineKac}

For constants $x_1$ and $x_2$, we note the simple limit formula 
\be
 \lim_{n,m,n'\to\infty}\vartheta_{n,m}(q,z^{\frac{1}{n'}})
 =\lim_{n,m,n'\to\infty}\sum_{\ell\in\mathbb{Z}}q^{m\ell^2+n\ell+\frac{n^2}{4m}}
   \,z^{\frac{m\ell}{n'}+\frac{n}{2n'}}
 =q^{x_1}z^{x_2}
\ee
where the limit is taken such that 
\be
 \lim_{n,m\to\infty}\frac{n^2}{4m}=x_1,\qquad \lim_{n,n'\to\infty}\frac{n}{2n'}=x_2
\ee
It follows that the logarithmic limit of the admissible characters (\ref{chadm}) is given by
\be
 \lim_{\hat{\rho},\hat{\rho}'\to\infty\atop \hat{\rho}/\hat{\rho}'\to \ph/\ph'}\chh_{r,s}^{\hat{\rho},\hat{\rho}'}\!(q,z) =\chih_{r,s}^{\,\ph,\ph'}(q,z)
\ee
where the limit is subject to
\be 
   \mathrm{gcd}(\ph,\ph')=1,\qquad \ph,\ph'\in\mathbb{N}
\label{pppplimits}
\ee
The resulting character
\be
  \chih_{r,s}^{\,\ph,\ph'}(q,z)
  =\frac{q^{h_{r,s}-\frac{c_k}{24}+\frac{1}{8}}\,z^{j_{r,s}+\frac{1}{2}}}{\eta(q,z)}
     \big(1-q^{rs}z^{-r}\big),\qquad r,s\in\mathbb{N}
\label{affKac}
\ee
is expressed here in terms of the central charge (\ref{ck}) and weights $j_{r,s}$ and $h_{r,s}$ given by
\be
 2j_{r,s}+1=r-st,\qquad h_{r,s}=\frac{(r-s t)^2-1}{4t},\qquad t=k+2=\frac{\ph}{\ph'}
\ee
We shall refer to these characters as {\em affine Kac characters}, and it is stressed that the corresponding
affine Kac labels $r$ and $s$ are 
unbounded from above. This is in stark contrast to the situation for the labels of admissible weights (\ref{j}),
but similar to the labelling of the Kac characters in (\ref{Kacchar}).
A detailed discussion of these characters and the underlying {\em affine Kac modules} will
appear in~\cite{PR2013b}.

\subsection{Logarithmic branching rules and branching functions}

A key result of this work is the determination of the {\em logarithmic branching functions} arising in the logarithmic
limit of the minimal branching rules (\ref{branchR}).
The corresponding central charges are
\be
 c^{p,p'\!;n}=\frac{3n}{n+2}\Big(1-\frac{2(n+2)(p'-p)^2}{n^2pp'}\Big)
  =\frac{3n}{n+2}-\frac{6(p'-p)^2}{npp'},\qquad 1\le p<p'
\label{cnpp}
\ee
where the integers $n,p,p'$ parameterize the logarithmic coset construction as in (\ref{Clog}). 
Curiously, we observe that for each $n$ there is exactly one logarithmic coset model with $c=0$
\be
 {\cal LM}(2,n+2;n):\qquad c^{2,n+2;n}=0,\qquad n\in\mathbb{N}
\ee
The first two members of this infinite sequence are critical percolation ${\cal LM}(2,3;1)$ and superconformal percolation ${\cal LM}(2,4;2)$~\cite{PRT}.

The logarithmic branching functions are obtained by taking the logarithmic limit of (\ref{b}) directly, subject to (\ref{pplimits}), 
\begin{align}
\chit_{r,s;\ell}^{\,p,p'\!;n}(q)&= \lim_{M,M'\to\infty\atop M/M'\to p/p'}b_{r,s;\ell}^{M,M'\!;n}(q)=q^{\Delta_{r,s}^{p,p'\!;n}-\frac{c^{p,p'\!;n}}{24}+\frac{n-1}{12(n+2)}}\bigg[
   \!\!\!\!\!\!\sum_{{0\le m\le n/2\atop m\equiveq\ell/2\;\text{mod 1}}\atop m\equiveq\pm(r-s)/2\; \mathrm{mod}\; n} \!\!\!\!\!c_{2m}^\ell(q) 
-\!\!\!\!\!\!\sum_{{0\le m\le n/2\atop m\equiveq\ell/2\;\text{mod 1}}\atop m\equiveq\pm(r+s)/2\; \mathrm{mod}\; n}
      \!\!\!\!\!q^{\frac{rs}{n}}\,c_{2m}^\ell(q) \bigg]
  \label{logBranch}    
      \\[-8pt]
 &=q^{\Delta_{r,s}^{p,p'\!;n}-\frac{c^{p,p'\!;n}}{24}+\frac{n-1}{12(n+2)}}
  \big[c_{r-s}^{\ell}(q)-q^{\frac{rs}{n}}c_{r+s}^{\ell}(q)\big],\quad
   r+s\equiveq \ell\ \mathrm{mod}\ 2,\quad
  r,s\in\mathbb{N},\quad 0\leq\ell\leq n
\nonumber
\end{align}
By definition of the string functions, these branching functions vanish if $r+s+\ell$ is odd. 
Explicitly, taking the logarithmic limit of (\ref{cosetDelta}), or by an analysis of (\ref{logBranch}), 
the logarithmic coset conformal weights are given by
\be
 \D_{r,s;\ell}^{p,p'\!;n}= \D_{r,s}^{p,p'\!;n}+\D_{r-s}^{\ell;n}+\mbox{Max}[\half(\ell\!+\!2\!-\!r\!-\!s),0],\qquad
 r+s\equiveq \ell\ \mathrm{mod}\ 2,\qquad r,s=1,2,\ldots
 \label{genLogConfWts}
\ee
The first term on the right side is given by (\ref{DrsMMn}) with the range of the replaced arguments $p,p'$ extended as in
(\ref{Clog}) while the Kac labels $r,s\in\mathbb{N}$ become unbounded. 
The second term is given by (\ref{stringConfWts}). The third term  
only gives a nonzero contribution for $r+s\le\ell\le n$. 
These conformal weights are thus organized into $n+1$ layered and {\em infinitely extended} Kac tables each displaying the 
checkerboard pattern (\ref{checkerboard}). In accord with the fact that these theories are non-unitary, the minimal conformal weight is
\be
\Delta^{p,p';n}_{\mathrm{min}}=\begin{cases}\Delta^{p,p';n}_{np,np';0},&\mbox{$p'-p$ even}\\[8pt]
\Delta^{p,p';n}_{np,np';n},&\mbox{$p'-p$ odd}
\end{cases}=\Delta^{p,p';n}_{np,np'}=-\frac{(p'-p)^2}{4n pp'}<0
\ee
It follows that the effective central charge is independent of $p,p'$ and given by the central charge of the affine current algebra $(A_1^{(1)})_n$
\be
 c_{\mathrm{eff}}^{\,p,p'\!;n}=c^{\,p,p'\!;n}-24\,\Delta^{p,p';n}_{\mathrm{min}}=\frac{3n}{n+2}=c_n
\ee
 
As discussed in~\cite{DiFMS}, for example, for small $n$, the string functions 
appearing in the branching functions can alternatively be written in terms of more familiar Virasoro minimal characters. 
For $n=1$, there is only one independent string function
\be
 c_0^0(q)=c_1^1(q)=\frac{1}{(q)_\infty}
\label{cn1}
\ee
It readily follows that the logarithmic branching functions (\ref{logBranch}) can be identified with the Kac characters (\ref{Kacchar}), that is,
\be
 \chit_{r,s;\ell}^{p,p'\!;n=1}(q)=\chit_{r,s}^{p,p'}(q),\qquad \ell= r+s\ \mathrm{mod}\ 2
\ee
since $\ell$ is uniquely determined. For $n=2$, there are three independent string functions which are related
to the three irreducible Virasoro characters $\ch_\D(q)$, $\D\in\{0,\frac{1}{16},\frac{1}{2}\}$, of the rational Ising model with central 
charge $c=c^{3,4;1}=\frac{1}{2}$ by
\be
c_0^0(q)
=c_2^2(q)
=\frac{\ch_{0}(q)}{(q)_\infty} ,\qquad
c_1^1(q)=\frac{\ch_{\frac{1}{16}}(q)}{(q)_\infty} ,\qquad
c_0^2(q)=\frac{\ch_{\frac{1}{2}}(q)}{(q)_\infty} 
\label{Ising}
\ee
The simplified expressions for the logarithmic branching functions that follow will be discussed in~\cite{PRT,PR2013b}.
For $n=3$, the four independent string functions are related
to irreducible Virasoro characters of the rational 3-state Potts model with central charge $c=c^{5,6;1}=\frac{4}{5}$ by
\be
c_0^0(q)=\frac{\ch_{0}(q)+\ch_{3}(q)}{(q)_\infty} ,\quad
c_1^1(q)=\frac{\ch_{\frac{1}{15}}(q)}{(q)_\infty} ,\quad
c_0^2(q)=\frac{\ch_{\frac{2}{5}}(q)+\ch_{\frac{7}{5}}(q)}{(q)_\infty} ,\quad
c_1^3(q)=\frac{\ch_{\frac{2}{3}}(q)}{(q)_\infty} 
\label{Potts}
\ee
Simplifications of the logarithmic branching functions for $n=3$ likewise follow from these expressions.

Finally, the logarithmic limit of the minimal branching rules 
(\ref{branchR}) follows from
\be
 \lim_{\hat{\rho},\hat{\rho}'\to\infty\atop \hat{\rho}/\hat{\rho}'\to \ph/\ph'}\chh_{r,s}^{\hat{\rho},\hat{\rho}'}(q,z)\,\chh_{\rho,0}^{n+2,1}(q,z)
  = \lim_{\hat{\rho},\hat{\rho}'\to\infty\atop \hat{\rho}/\hat{\rho}'\to \ph/\ph'}
  \sum_{\mbox{\scriptsize$\sigma=1$}\atop \mbox{\scriptsize $\sigma\equiveq r\!+\!\ell$ mod 2}}^{\hat{\rho}+n\hat{\rho}'-1}
    b_{r,\sigma;\ell}^{\,\hat{\rho},\hat{\rho}+n\hat{\rho}'\!;n}(q)\,\chh_{\sigma,s}^{\hat{\rho}+n\hat{\rho}'\!,\hat{\rho}'}\!(q,z)
\label{branchL1}
\ee
and yields the {\em logarithmic branching rules}
\be
 \chih_{r,s}^{\,\ph,\ph'}(q,z)\,\chh_{\rho,0}^{n+2,1}(q,z)
  =\sum_{\mbox{\scriptsize$\sigma\in\mathbb{N}$}\atop \mbox{\scriptsize $\sigma\equiveq r\!+\!\ell$ mod 2}}
    \chit_{r,\sigma;\ell}^{\,\ph,\ph+n\ph'\!;n}(q)\,\chih_{\sigma,s}^{\,\ph+n\ph'\!,\ph'}\!(q,z)
\label{branchL2}
\ee 
In~\cite{PR2013b}, we present an independent algebraic analysis of the multiplication formula (\ref{branchL2}) from a constructive approach using representation theory.

\section{Conclusion}
\label{SecConcl}

In this paper, we have generalized the logarithmic minimal models ${\cal LM}(p,p')={\cal LM}(p,p';1)$~\cite{PRZ0607} to the higher fusion level logarithmic minimal models ${\cal LM}(p,p';n)$ with $n\in\mathbb{N}$. 
From the lattice~\cite{PRT}, these new models are constructed by fusing together $n\times n$ blocks of the elementary face weights of appropriate ${\cal LM}(p,p')$ models. The associated CFTs are expected to be logarithmic in the sense that they admit indecomposable representations of rank greater than $1$. 
We argue that the complete family of associated logarithmic 
CFTs  ${\cal LM}(p,p';n)$ can be realized by a {\em Virasoro coset construction} (\ref{Clog}) in analogy to the fractional level $su(2)$ coset construction~\cite{GKO85,GKO86,ACT91} of the minimal models ${\cal M}(p,p';n)$.
By applying a {\em logarithmic limit}~\cite{Ras0405,Ras0406} to the known minimal branching rules, we obtain explicit formulas for the {\em logarithmic branching rules} (\ref{branchL2}) expressing the decomposition of the relevant affine characters. 
The {logarithmic branching functions} $\chit_{r,s;\ell}^{\,p,p'\!;n}(q)$, which appear as coefficients in these branching rules and are given explicitly by (\ref{logBranch}), play the role of {\em Kac characters}~\cite{PRZ0607,Ras1012,BGT1102} for the logarithmic $su(2)$ coset theories. 
Our results constitute the first logarithmic coset construction in the Virasoro picture, without resource to a ${\cal W}$-symmetry.
Algebraic proofs of some of the ensuing character relations will be presented in~\cite{PR2013b}.
There we also discuss the structure of the {\em affine Kac modules} associated with the affine Kac characters. A further analysis of the representation theory of the logarithmic coset models will rely on 
the detailed representation theory of affine current algebras at fractional level including the Jordan-block structures of higher-rank
indecomposable representations~\cite{Gab0105,LMRS0311,CR1205}, see also~\cite{KN0112,Nic0210,Ras0508}.

We stress that our logarithmic coset construction and logarithmic branching functions are completely general applying to the ${\cal LM}(p,p';n)$ models for all integer fusion levels  
$n\in\mathbb{N}$ and all fractional levels $k=\frac{np}{p'-p}-2$ with $\mbox{gcd}(p,\frac{p'-p}{n})=1$, $1\le p<p'$ 
and $p,p'\in \mathbb N$. The $su(2)$ coset realization of these theories takes us some way towards fitting logarithmic 
theories~\cite{FT1002} into the usual Lie algebraic classification schemes that are applied to rational theories. 

This paper opens several avenues for further research. Since they represent building blocks, it would be of interest to systematically study the affine logarithmic models associated with $(A_1^{(1)})_k$ directly from the lattice. Clearly, it is of interest to extend the coset construction of the logarithmic $su(2)$ models to the 
${\cal W}$-extended picture and to obtain in this way general expressions for the ${\cal W}$ branching functions.
Without employing coset constructions, extensions of fractional-level $su(2)$ current algebras to logarithmic 
${\cal W}$-algebra structures have recently been obtained in~\cite{ST1301}. 
Since fractional levels play a key role, a related question is to see if a logarithmic coset construction makes sense when both $n$ and $k$ are fractional levels. 
Coset constructions of rational models with ${\cal W}$-extended symmetries based on higher-rank 
$su(N)$ Lie algebras with $N>2$ were studied in~\cite{FL1988,BBSS1988,BS9210}.
It would also be worthwhile to see if similar coset constructions exist for logarithmic theories based~\cite{Shigechi} 
on higher-rank $su(N)$ Lie algebras. 

While completing the writing up of the present work, the preprint~\cite{CRW1305} appeared in which a different logarithmic
coset construction is studied. It pertains to $n=1$ and $p=1$ and details 
of the corresponding branching functions are provided for $(p,p')=(1,2)$ and $(1,3)$ only, but unlike the present work, 
it also considers certain ${\cal W}$-extensions. It is not immediately clear if the construction in~\cite{CRW1305}
is related to ours.

\section*{Acknowledgments}
\vskip.1cm
\noindent
JR is supported by the Australian Research Council (ARC) under the ARC Future Fellowship scheme, project number FT100100774. 
We thank Changrim Ahn for helpful discussions on fractional level branching rules and Alexi Morin-Duchesne for reading and
commenting on the manuscript.



\end{document}